\documentclass[12pt]{article}
\usepackage{graphicx}
\begin{document}
\title{Why Solve the Hamiltonian Constraint in Numerical Relativity?} 
\author{Beverly K. Berger\\
{\small Physics Division, National Science Foundation, Arlington, VA 22207 USA}}
\date{}
\maketitle

\begin{abstract}
The indefinite sign of the Hamiltonian constraint means that solutions to Einstein's equations must achieve a delicate balance---often among numerically large terms that nearly cancel. If numerical errors cause a violation of the Hamiltonian constraint, the failure of the delicate balance could lead to qualitatively wrong behavior rather than just decreased accuracy. This issue is different from instabilities caused by constraint-violating modes. Examples of stable numerical simulations of collapsing cosmological spacetimes exhibiting local mixmaster dynamics with and without Hamiltonian constraint enforcement are presented.  
\end{abstract}
\section{Introduction \protect \footnote{This paper is based on a talk given at the 17th International Conference on General Relativity and Gravitation.}}
It gives me great pleasure to dedicate this note to Mike Ryan in honor of his 60th birthday. It is especially rewarding to be able to include several of Mike's favorite research themes including Mixmaster dynamics and spatially inhomogeneous cosmologies---where he has made important contributions.

The $3+1$ formulation of Einstein's equations has served as the starting point for most numerical simulations. While the precise set of Einstein's equations most suitable for simulations is a subject of active investigation (for a review see \cite{baumgarte03}), the basic structure of the ADM form---evolution equations for (something related to) the induced spatial metric and (something related to) the spatial extrinsic curvature plus (some form of) the Hamiltonian and momentum constraints---will most likely remain. As is well known, the Einstein evolution equations preserve the constraints but the discretized evolution equations need not do so. To date, most simulations of binary black holes and/or neutron stars have solved only the evolution equations. Experience with these simulations has demonstrated that failure to solve the constraints allows the growth of constraint violating modes that can cause the codes to crash. The main argument against numerical enforcement of the constraints has been the cost in compute time of solving the elliptic constraint equations. Advances in computational power, improved elliptic solvers, and the growing awareness of the problem of instabilities caused by constraint violation have led to increased interest in constrained evolution \cite{calabrese04}.

In this note, I shall discuss a further reason to consider constrained evolution---especially for the Hamiltonian constraint. The Hamiltonian constraint is given schematically by
\begin{equation}
\label{hamgen}
{\cal H}^0 = {1 \over {\sqrt{h}}} \left( K^{ij}K_{ij}\,-\,K^2 \right) \,-\, \sqrt{h} {}^3R[h_{ij}] = 0
\end{equation}
where $K_{ij}$ and $h_{ij}$ are respectively the extrinsic curvature and induced metric of the spatial hypersurface, $h$ is the determinant of $h_{ij}$,  ${}^3R[h_{ij}]$ is the scalar curvature of $h_{ij}$, and $K$ is the trace of $K_{ij}$. Note first the indefinite sign of ${\cal H}^0$ which can appear in either the ``kinetic'' terms involving $K_{ij}$ or in the ``potential" term containing  ${}^3R$. Thus to maintain the solution ${\cal H}^0 = 0$ requires a delicate balance among the variables. Failure to achieve this balance in a numerical simulation can yield not only numerical instability but also---even in a stable evolution---qualitatively incorrect behavior. In the following, three examples of collapsing cosmological spacetimes with 3, 2, and 1 spatial Killing field will be used to illustrate qualitatively incorrect behavior that results if the Hamiltonian constraint is allowed to evolve freely in a simulation.

Collapsing cosmological spacetimes are characterized by two main types of behavior---asymptotic velocity term dominance (AVTD) and local Mixmaster dynamics (LMD)---first described by Belinskii et al (BKL) (e.g.\ \cite{belinskii71b}). Both types of behavior arise when the dynamics becomes local---variables at each spatial point evolve as a separate spatially homogeneous universe. The basic building block of the approach to the singularity is the Kasner universe, characterized by fixed collapse rates along the principle spatial axes. This is in contrast to Bianchi IX (Mixmaster) collapse where an infinite sequence of ``bounces'' off the spatial scalar curvature change one Kasner epoch into another with different fixed collapse rates. The relationship between one Kasner epoch and the next includes bounces that are sensitive to initial conditions and can be determined from conservation of ``momentum'' through the bounce \cite{belinskii71b,misner69}.

Collapsing cosmologies exhibit LMD if, at (almost)\footnote{The word ``almost'' will be used in this context to indicate that set-of-measure-zero exceptions are known. Details of these can usually be found in the cited references.} every spatial point, bounces from one Kasner-like epoch to another may be demonstrated and shown to obey the relevant bounce laws. In contrast, AVTD behavior is characterized by a final Kasner epoch at (almost) every spatial point after a possible final bounce. Rigorous demonstration of AVTD behavior has been provided \cite{andersson00}. With the exception of spatially homogeneous cosmologies \cite{ringstrom99}, LMD has been plausibly demonstrated only through numerical simulation \cite{berger98c,berger01,garfinkle03}.

In the remainder of this note, we shall consider three examples where LMD can be seen only if the Hamiltonian constraint is explicitly enforced. These examples are the spatially homogeneous vacuum, diagonal Bianchi IX cosmology \cite{belinskii71b,misner69,berger96c}, the two spatial Killing field generic $T^2$-symmetric vacuum spacetime \cite{berger97g,berger01}, and the one spatial Killing field vacuum $U(1)$-symmetric cosmology \cite{moncrief86,berger98a}. While failure to enforce the Hamiltonian constraint can lead to instability, we shall focus on simulations that evolve stably but indicate that the approach to the singularity is AVTD rather than LMD. 

\section{Examples}
\subsection{Bianchi IX (Mixmaster) Cosmology}
The Mixmaster universe is described by the metric \cite{misner69}
\begin{equation}
\label{mixmastermetric}
ds^2 = - e^{3\Omega} d\tau^2 + e^{2\Omega} \left(e^{2\beta}\right)_{ij} d\sigma^id\sigma^j
\end{equation}
where $\beta_{ij} = {\rm diag} (-2\beta_+,\,\beta_++2 \sqrt{3}\beta_-,,\beta_+-2 \sqrt{3}\beta_-)$, $\Omega$, $\beta_\pm$ depend only on the time $\tau$, and the spatial 1-forms $\sigma^i$ satisfy the appropriate $SU(2)$ relationship for Bianchi IX. Einstein's equations may be found from the variation of the superhamiltonian (lapse times the Hamiltonian constraint) specialized for these models as \cite{misner69}
\begin{equation}
\label{mixmasterH}
2H^0 = 0 = -\,p_\Omega^2 \,+\,p_+^2 \,+\,p_-^2\,+\,V(\Omega,\beta_\pm)
\end{equation}
where $p_\Omega$, $p_\pm$ are canonically conjugate to $\Omega$,$\beta_\pm$ and
\begin{equation}
\label{mixmasterV}
V(\Omega, \beta_\pm) = e^{4\Omega-8\beta_+}\,+\,e^{4\Omega +4 \beta_+ +4\sqrt{3}\beta_-}\,+\,e^{4\Omega +4 \beta_- +4\sqrt{3}\beta_-} \,+ \,\ldots
\end{equation}
where the ellipsis indicates terms that are (almost) always negligible.
In the absence of the potential $V$ (with $\sigma^i = dx^i$), the Kasner solution is obtained. Eq.\ (\ref{mixmasterH}) defines the dynamics in minisuperspace (MSS). The Kasner solution represents the free particle in MSS. For the Kasner solution, Eq.\ (\ref{mixmasterH}) may be written as 
\begin{equation}
\label{kasnercircle}
K \equiv v_+^2\,+\,v_-^2 = 1
\end{equation}
where $v_\pm = -p_\pm/p_\Omega$. The addition of the potential (\ref{mixmasterV}) causes (almost) every Kasner epoch to end in a bounce off one of the exponential terms in $V$. After the bounce, the behavior is again described by (\ref{kasnercircle}) but with different Kasner parameters. As first discussed by BKL, every Kasner epoch can be identiifed by a single parameter $u$ (related to the anisotropic collapse rates) such that the $n+1^{st}$ Kasner epoch is related to the $n^{th}$ one in Mixmaster dynamics through
\begin{equation}
\label{umap}
u_{n+1} = \left\{ {\matrix{{u_n - 1} \quad & {u_n \ge 2} \cr
{1 \over {u_n - 1}} \quad & 1 \le u_n \le 2}} \right.\ .
\end{equation}
As the Mixmaster universe evolves toward the singularity, the ratio of the duration of the bounce to the duration of the Kasner epoch goes to zero. The role of the Hamiltonian constraint (\ref{mixmasterH})--- as it becomes arbitrarily close to (\ref{kasnercircle}) except at the bounce---is to keep the configuration on the ``Kasner circle'' defined by $K = 1$. A typical evolution may be found in \cite{bkblr} , Fig.\ 5.

\begin{figure}
\center
\includegraphics[scale = .45]{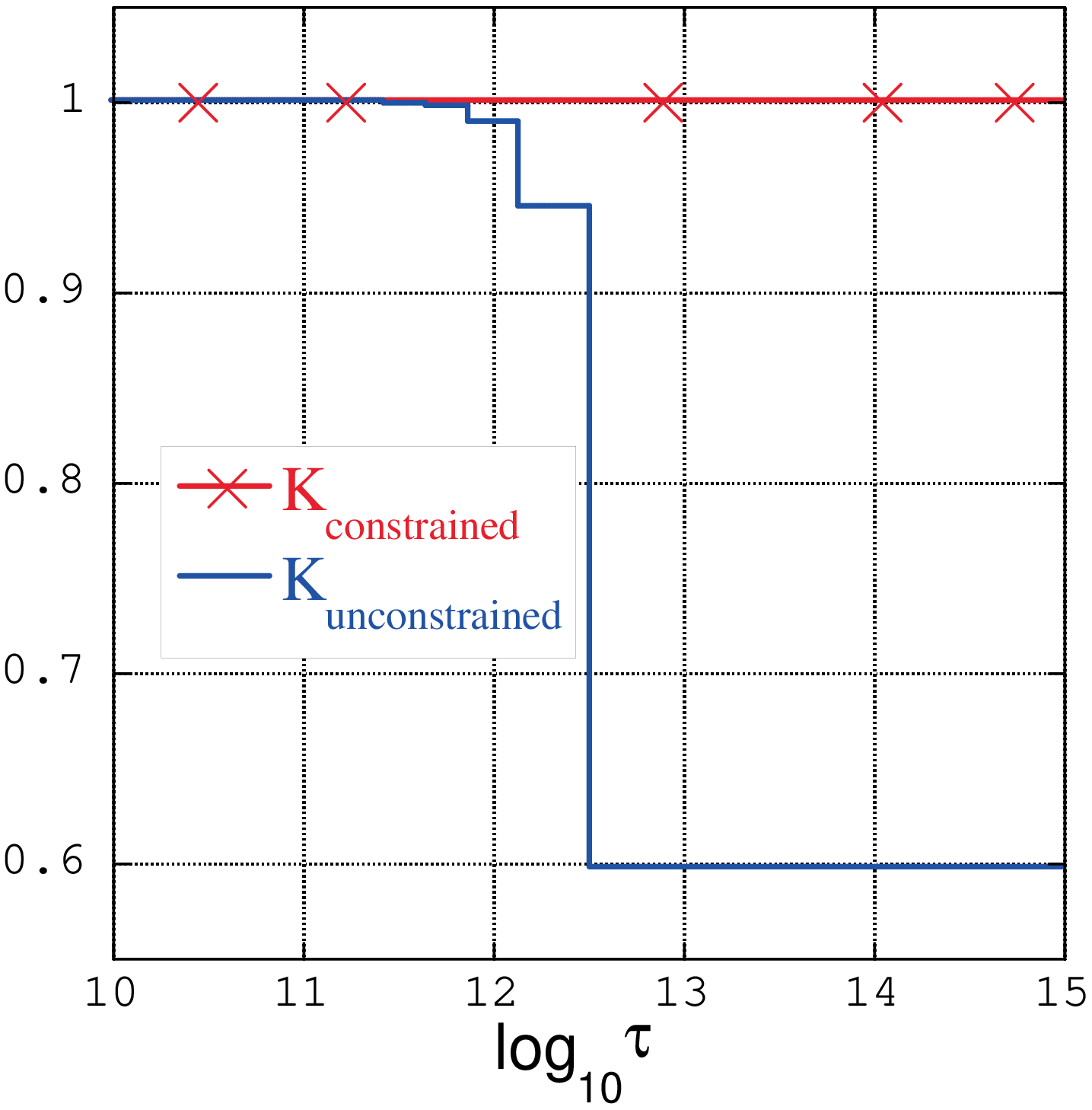}
\includegraphics[scale = .45]{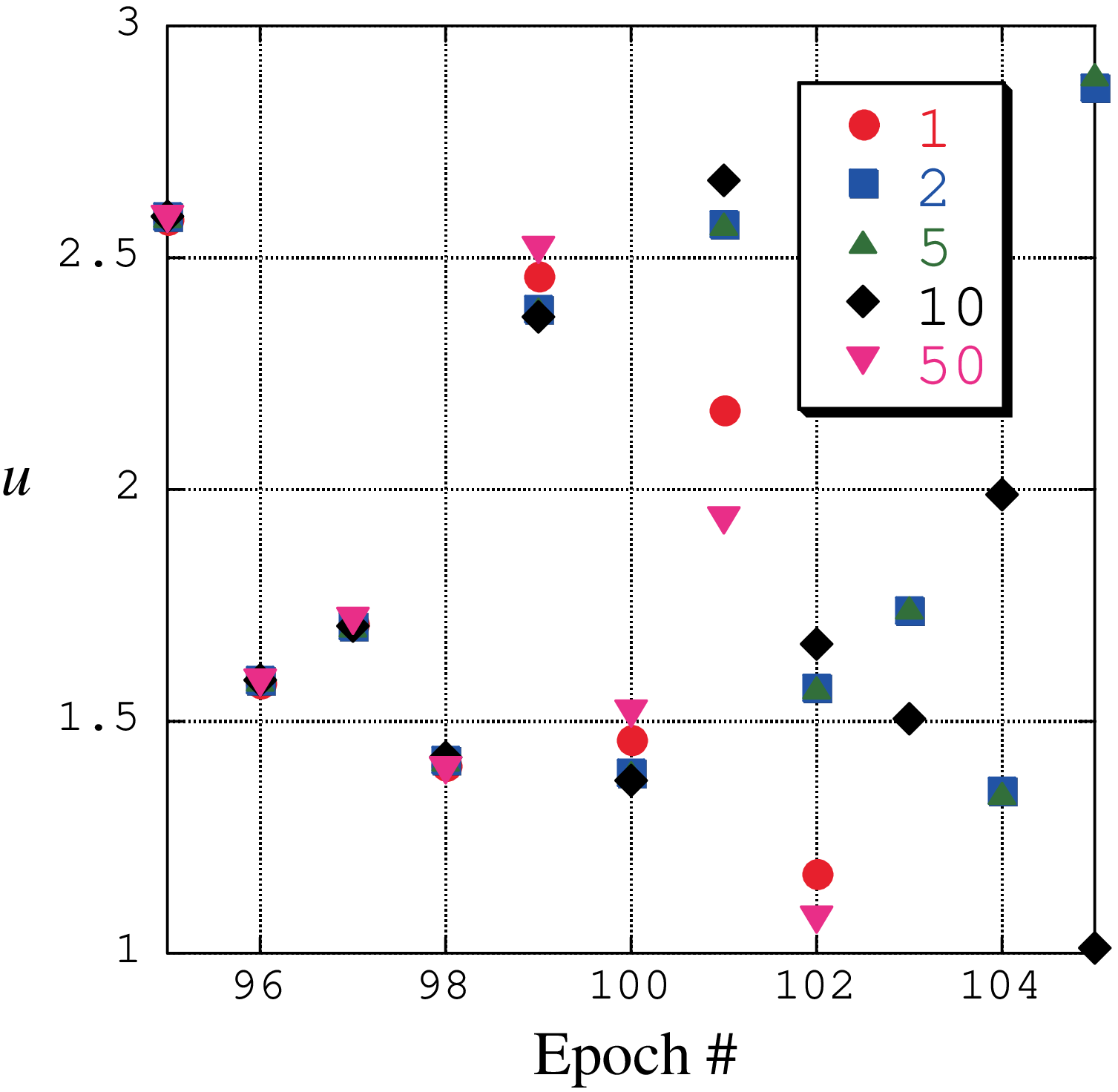}
 \caption{\label{mixkasnercircle} \small{ Constraint enforcement in collapsing Bianchi IX cosmologies. Left: Comparison of constrained and unconstrained evolutions showing $K$ vs $\tau$. Right: Sequence of $u$-values for constraint enforcement every $N$ time steps. }}
\end{figure}

A numerical simulation of Mixmaster collapse with a code that can follow the evolution through hundreds of bounces \cite{berger96c} shows qualitatively different behavior depending upon whether or not the Hamiltonian constraint is enforced. Figure \ref{mixkasnercircle} shows a comparison of the Kasner circle indicator $K$ vs time $\tau$ for constrained and unconstrained simulations. The constrained simulation maintains $K = 1$ (except at bounces) while the unconstrained one evolves to $K < 1$. With $K < 1$, the system point is moving too slowly in MSS to bounce off the potential. Thus the unconstrained simulation yields the spurious result that the model has a last bounce and is AVTD. The simulation does not become unstable. There is no code crash---just a physically wrong result. (From some initial data, the unconstrained evolution leads to $K > 1$. This does become unstable and crashes.) Figure \ref{mixkasnercircle} also compares the evolution from the same initial data while solving the constraint at every $N$ time steps for $N = 1,\ 2,\ 5,\ 10,\ 50$. The evolution is monitored by the sequence of $u$-values obtained for the Kasner segments. Two evolutions become qualitatively distinct when the integer parts of $u$ differ for the same segment. The sequence of $u$-values obtained from (\ref{umap}) is sensitive to initial conditions when an era ends [$u \to 1/(u-1)$]. Note that the evolutions are qualitatively identical for a large number of bounces independent of $N$ and that they begin to differ at about the same place. This epoch is essentially the point at which information about the initial value of $u$ is lost (in the double precision that is used). Thus one cannot really choose one of the solutions as more correct than another. The lesson here appears to be that constraint solving---at least from time to time---is essential for the accurate simulation of collapsing homogeneous Mixmaster models.

\subsection{$U(1)$-Symmetric Cosmology}
As our next example, we consider the vacuum $U(1)$-symmetric cosmologies discussed extensively by Moncrief \cite{moncrief86}. The metric for $T^3$ spatial topology is given by (for a specific choice of lapse and shift)
\begin{equation}
\label{u1metrica}
ds^2 = e^{-2 \varphi} \left[-  e^{2\Lambda} \, d\tau^2 + e^{\Lambda}\,e_{ab}(x,z) d\xi^a d\xi^b \right] 
+ e^{2 \varphi} (d\xi^3+ \beta_a
\, dx^a  \, d\tau)^2
\end{equation}
where $a,\,b = 1,\ 2$ and $\varphi,\ \Lambda, \ x,\ z,$ and $\beta_a$ depend on spatial variables $\xi_1,\ \xi_2,$ and time $\tau$. The explicit form of $e_{ab}$ is given in \cite{moncrief86,berger97e} as is the discussion of a canonical transformation to replace the twists $\beta_a$ with a single twist potential $\omega$. Einstein's equations are obtained from variation of the Hamiltonian \cite{berger98a}
\begin{eqnarray}
\label{Hu1}
H &=& \int_{T^3} \left[ \left( {1 \over 8}p_z^2+{1 \over 2}
e^{4z}p_x^2+{1 \over 8}p^2+{1 \over 2}e^{4\varphi }r^2-{1 \over 2}p_\Lambda
^2  \right) \right. \nonumber \\
&& + \left\{  \left( {e^\Lambda e^{ab}} \right) ,_{ab}-
\left( {e^\Lambda e^{ab}}
\right) ,_a\Lambda ,_b+e^\Lambda  \right. \left[  \left( {e^{-2z}}
\right) ,_u x,_v- \left( {e^{-2z}} \right) ,_v x,_u \right] \nonumber \\
&& \left. \left. +2e^\Lambda e^{ab}\varphi ,_a\varphi ,_b+{1 \over 2}
e^\Lambda e^{-4\varphi }e^{ab}\omega ,_a\omega ,_b \right\} \right] = \int_{T^3} \,{\cal H}
\end{eqnarray}
where ${\cal H}= 0$ is the Hamiltonian constraint and $p_\varphi$, $r$, $p_\Lambda$, $p_z$, and $p_x$ are cannonically conjugate to $\varphi$, $\omega$, $\Lambda$, $z$, and $x$. 

\begin{figure}
\center
\includegraphics[scale = .45]{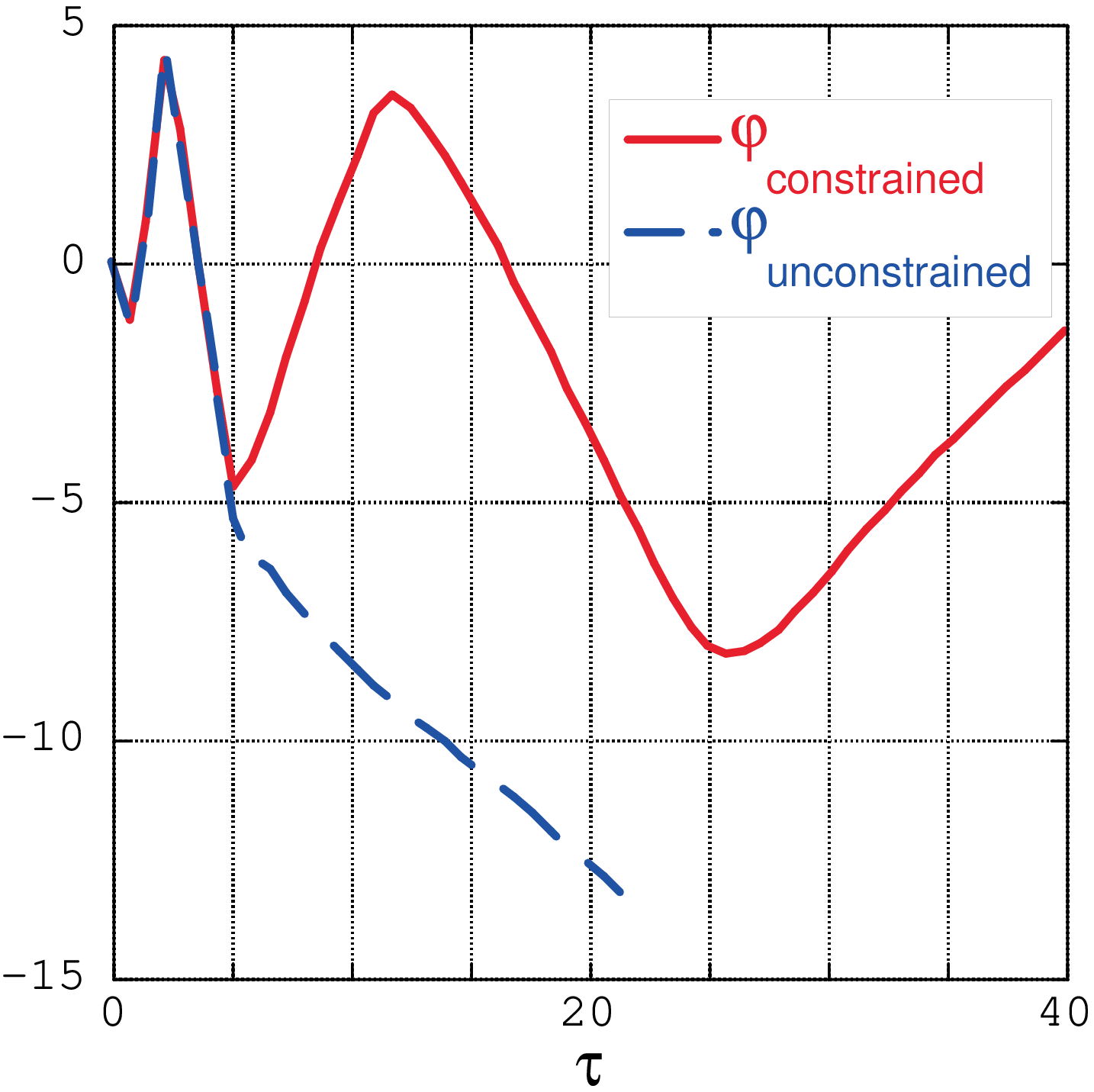}
\includegraphics[scale = .45]{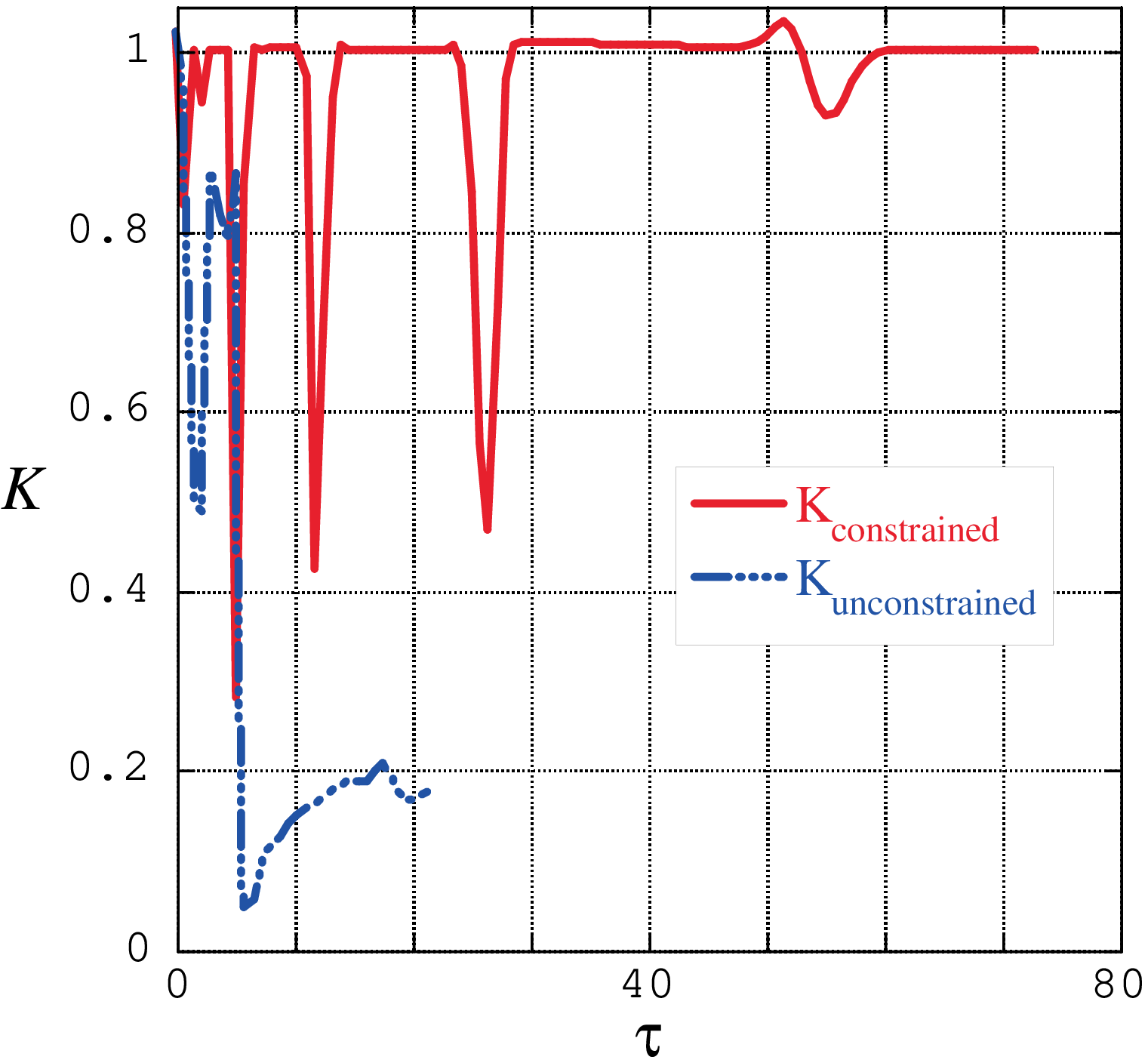}
 \caption{\label{u1phi} \small{ Comparison of $\varphi$ vs $\tau$ (left) and $K$ vs $\tau$ (right) for constrained and unconstrained evolutions in $U(1)$-symmetric collapse from the same initial data. Behavior at a representative spatial point is shown. }}
\end{figure}

Polarized $U(1)$-symmetric models ($\omega = 0 = r$) have been examined both numerically \cite{berger97e} and analytically \cite{isenberg02} where the case has been made that the singularity is AVTD. Generic vacuum models are, so far, beyond analytic study. Numerical simulations indicate that $U(1)$-symmetric collapse exhibits LMD \cite{berger98a,hern00}. This behavior can be understoon in terms of the Method of Consistent Potentials (MCP) \cite{berger98c} or by analogy with spatially homogeneous models \cite{berger99b}. Snapshots from a typical evolution are shown in \cite{berger98a}. As was mentioned in \cite{berger98a}, the LMD behavior cannot be achieved without explicit enforcement of the Hamiltonian constraint. If the momentum constraint is not enforced, the Hamiltonian constraint may be solved algebraically (e.g.\ for $p_\Lambda$). Figure \ref{u1phi} illustrates that the wave amplitude $\varphi$ exhibits LMD oscillations at a representative spatial point in a constrained evolution while, from the same initial date at the same spatial point, the oscillations are absent in unconstrained evolution. Figure \ref{u1phi} also compares the analog of the Kasner measure $K$ (defined so that $K = (p/4p_\Lambda)^2+ (p_z/4p_\Lambda)^2 = 1$ is the analog of the Kasner circle) vs $\tau$ for the same simulations. Once again, there is no instability in either case. As in the spatially homogeneous example, failure to enforce the Hamiltonian constraint leads to qualitatively incorrect behavior---a spurious indication that the model is AVTD.

\subsection{Generic $T^2$-Symmetric Collapse}
Our final example will be generic $T^2$-symmetric collapse \cite{berger01}. These spacetimes are the most general $T^2$-symmetric vacuum spacetimes and reduce to the Gowdy model with $T^3$ spatial topology \cite{gowdy71} if the twists are set to zero. As in \cite{berger01}, consider the metric
\begin{eqnarray}
\label{galmetricnew}
ds^2&=&-e^{(\lambda -3\tau )/2}d\tau ^2+e^{(\lambda +\mu +\tau )/2}d\theta
^2+ e^{-P-\tau }[d\delta-(\int^\tau  d\tau'\,\Theta)\,d\theta]^2 \nonumber \\
&&+e^{P-\tau }\left\{ d\sigma+Qd\delta +\left[ \int^\tau d\tau'\, (Q \Theta) - Q \int^\tau d\tau'\, \Theta\right]\,d\theta\right\}^2
\end{eqnarray} 
where the wave amplitudes $P$ and $Q$ and the ``background'' $\lambda$ depend only on the spatial variable $\theta$ and time $\tau$, $\Theta = \kappa e^{(\lambda + 2P + 3 \tau)/2} e^{\mu/4} $ for $\kappa$ the twist constant, and $e^{\mu/4} = 2 \pi_\lambda$. (For details see \cite{berger01}.) Einstein's equations may be derived from the Hamiltonian density ({\it which is not the Hamiltonian constraint})
\begin{equation}
\label{hgal}
{\cal H} ={1 \over {4 \pi_\lambda}} \left[\pi_P^2 + e^{-2P} \pi_Q^2 + e^{-2\tau} \left( P,_\theta^2 + e^{2P} Q,_\theta^2 \right) \right] 
+ \sigma \, \kappa^2 \, \pi_\lambda
e^{(\lambda + 2P + 3\tau ) / 2} .
\end{equation}
This is supplemented by the momentum constraint $\pi_P P,_\theta + \pi_Q Q,_\theta + \pi_\lambda \lambda,_\theta = 0$. The Gowdy model is recovered if $\kappa = 0,\ \pi_\lambda = \textstyle{1\over 2}$.
The first order equation for $\lambda$ obtained from the variation of (\ref{hgal}) is in fact the Hamiltonian constraint. Thus these variables and equations ``automatically'' enforce the Hamiltonian constraint. A typical evolution is shown in \cite{berger01}.

\begin{figure}
\center
\includegraphics[scale = .45]{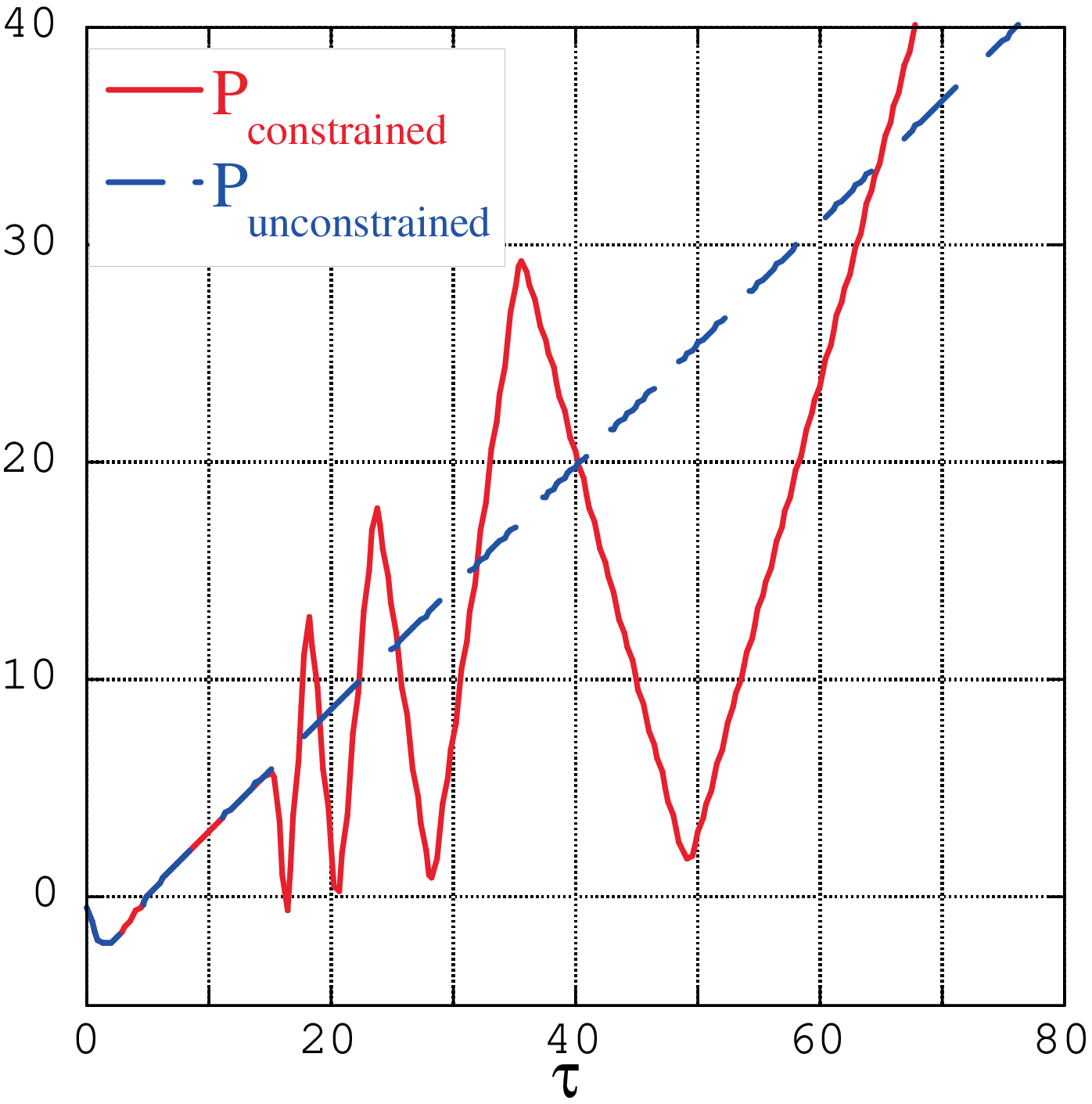}
\includegraphics[scale = .45]{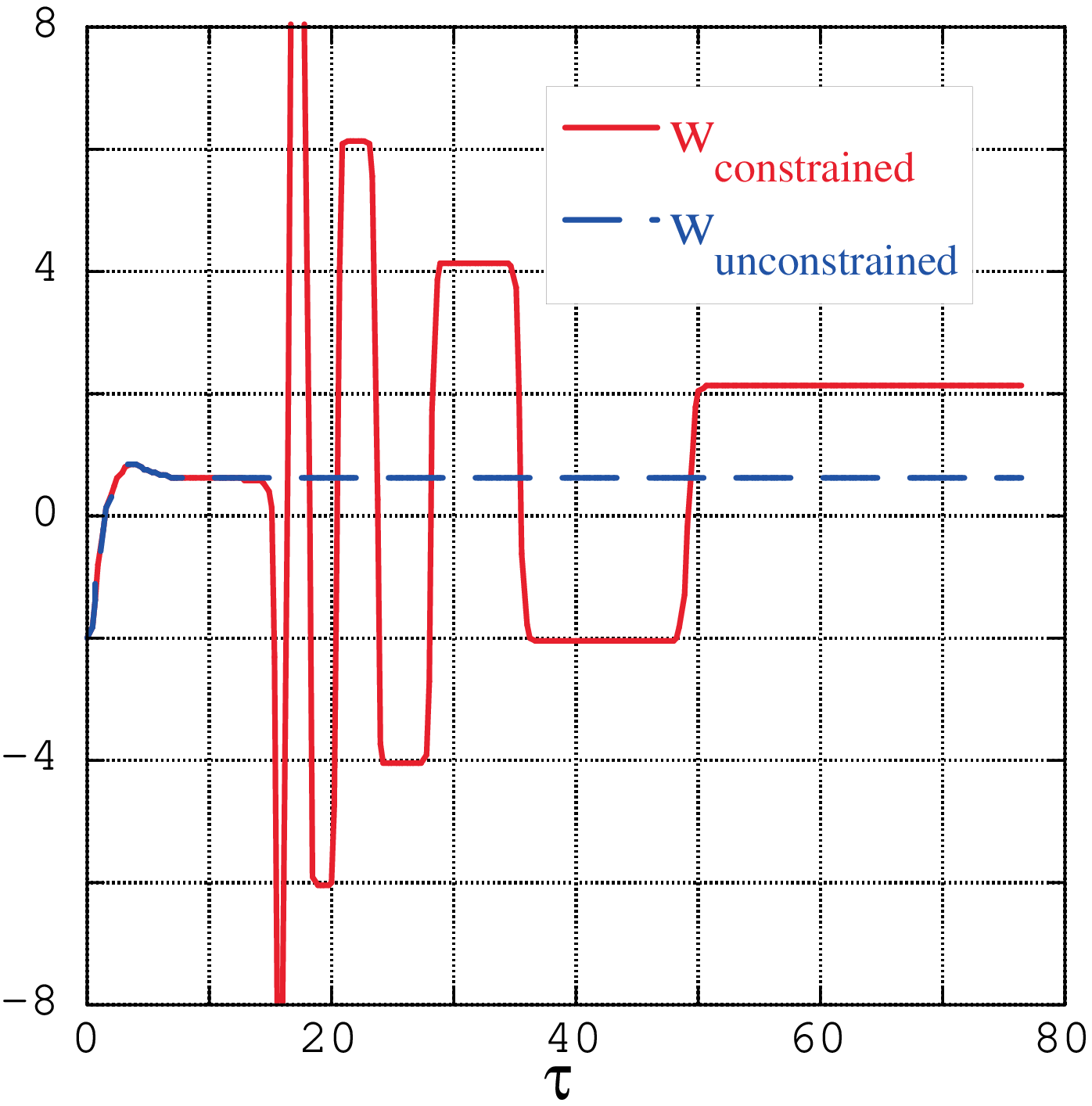}
 \caption{\label{galP} \small{ Comparison of $P$ vs $\tau$ (left) and $w$ vs $\tau$ (right) for constrained and unconstrained evolutions in generic $T^2$-symmetric collapse from the same initial data. Behavior at a representative spatial point is shown. }}
\end{figure}

The set of equations obtained from the variation of (\ref{hgal}) plus the momentum constraint do not comprise the full set of Einstein equations obtained from (\ref{galmetricnew}) and expressed as $G_{\mu\nu}=0$. One also obtains a redundant wave equation for $\lambda$ (see \cite{berger97g}). This wave equation may be used instead of the Hamiltonian constraint to evolve $\lambda$ to produce an unconstrained evolution. Figure \ref{galP} shows respectively the wave amplitude $P$ and $v$-like parameter $w = \pi_P/2\pi_\lambda$ (see \cite{berger01}) at the same representative spatial point for constrained and unconstrained evolution from the same initial data. It is clear that the LMD discussed in detail and understood quantitatively in \cite{berger01} through ``bounce laws'' for $w$ cannot be reproduced without explicit enforcement of the Hamiltonian constraint. Once again, there is no numerical instability associated with failure to enforce the constraint.

\section{Conclusions}
Three examples have been presented where failure to enforce the Hamiltonian constraint has led, not to instability, but to qualitatively incorrect behavior. The Hamiltonian constraint consists of kinetic energy-like and potential energy-like terms. In collapsing cosmologies with LMD behavior, the kinetic terms either dominate or intermittently dominate the dynamics. If the Hamiltonian constraint is not enforced, the relationship among the momenta in the kinetic term will become incorrect. However, these momenta comprise the coefficients $\alpha$ of the time $\tau$ in potential terms of the form $e^{\alpha \tau}$. How these terms grow and decay determine whether the approach to the singularity is AVTD or LMD (see the discussion in \cite{berger98c}). With the wrong coefficient---or even the wrong sign for $\alpha$---a term that should grow exponentially may decay and vice versa in a numerical simulation. 

There are collapsing cosmological spacetimes that do not require explicit enforcement of the Hamiltonian constraint. Examples already mentioned include the polarized $U(1)$ models and the Gowdy cosmologies. Thus it seems as if the ``delicate'' LMD behavior must be present for enforcement of the Hamiltonian constraint to be needed. Enforcement of the momentum constraint appears to be less critical. However, momentum constraint violation can be shown to lead to incorrect spatial waveforms in collapsing cosmologies. Garfinkle has demonstrated LMD in generic collapse without explicit constraint enforcement (although damping was introduced to suppress constraint violating modes) \cite{garfinkle03}. The variables he used are more naturally adapted to the constraint hypersurface but are not suitable for binary black hole evolution \cite{garfinkle04}.

The lesson from the examples presented here is that failure to enforce the Hamiltonian constraint can yield qualitatively incorrect behavior in a numerical simulation even in the absence of instability. While this does not occur in all classes of spacetimes and for all choices of variables and formulations, failure to enforce the constraint may be dangerous when properties of the solution are unknown.

\section*{Acknowledgment}
This research was supported by the National Science Foundation.

\end{document}